\documentclass[amssymb,amsmath,prl,twocolumn,floatfix,superscriptaddress,showpacs]{revtex4}
\usepackage{graphicx}
\usepackage{lastpage}
\usepackage{amssymb}
\usepackage{epsfig}
\newcommand{\ket}[1]{|#1 \rangle}

\begin{document}

\title{The Photonic Module: an on-demand resource for photonic entanglement.}
     \author{Simon J. Devitt}
    \email{devitt@physics.unimelb.edu.au}
    \address{Centre for Quantum Computing Technology, School of Physics,
    University of Melbourne, Victoria 3010, Australia.}
      \author{Andrew D. Greentree}
     \address{Centre for Quantum Computing Technology, School of Physics,
    University of Melbourne, Victoria 3010, Australia.}
    \author{Radu Ionicioiu}
    \address{Hewlett-Packard Laboratories, Filton Road, Stoke Gifford, Bristol BS34 8QZ, UK} 
    \author{Jeremy~L.~O'Brien}
        \address{Centre for Quantum Photonics, H. H. Wills Physics Laboratory \& Department of Electrical and Electronic Engineering, University of Bristol, Merchant Venturers Building, Woodland Road, Bristol, BS8 1UB, UK}
        \author{William J. Munro}
    \address{Hewlett-Packard Laboratories, Filton Road, Stoke Gifford, Bristol BS34 8QZ, UK} 
     \address{ 
National Institute of Informatics, 2-1-2 Hitotsubashi, Chiyoda-ku, Tokyo 101-8430, Japan} 
    \author{Lloyd C.L. Hollenberg}
     \address{Centre for Quantum Computing Technology, School of Physics,
    University of Melbourne, Victoria 3010, Australia.}

\date{\today{}}
\begin{abstract}
Photonic entanglement has a wide range of applications in quantum computation and communication. Here we introduce a new device: the Òphotonic moduleÓ, which allows for the rapid, deterministic preparation of a large class of entangled photon states. The module is an application independent, ``plug and play" device, with sufficient flexibility to prepare entanglement for all major quantum computation and communication applications in a completely deterministic fashion without number-discriminated photon detection. We present two alternative constructions for the module, one using free-space components and one in a photonic bandgap structures. The natural operation of the module is to generate states within the stabilizer formalism and we present an analysis on the cavity-QED requirements to experimentally realize this device.

\end{abstract}
\maketitle

\section{Introduction}

\begin{figure*}[tb!]
\includegraphics[width=1.6\columnwidth,clip]{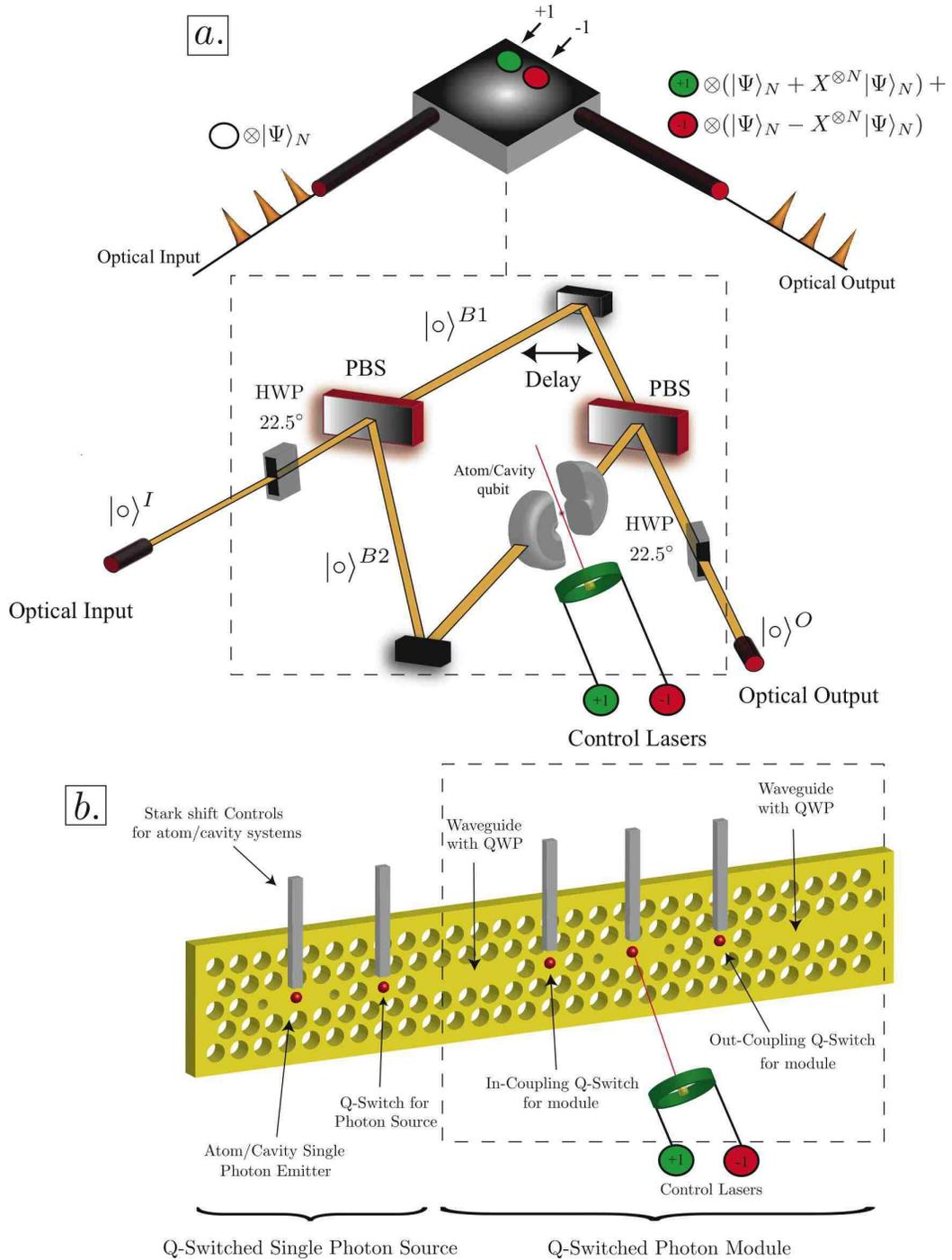}
\caption{\label{fig:Fig1} (Colour Online) 
\textbf{Schematics showing the basic design of a photonic module (components within the dashed boxes) in 
freespace and photonic crystals. a.} 
Photonic module design for a polarization independent atom/photon interaction, requiring two 
HWP and two PBS in free space.  The polarization dependent 
interferometer ensures that 
only the vertical component of the single photon interacts with the atom/cavity qubit.  
\textbf{b.} A photonic bandgap structure for a polarization dependent atom/photon version of 
the photonic module, required to prepare two photon Bell states and higher order GHZ states.  
The initial two cavities represent a Q-switched single photon source~\cite{Andy}.  Single photons are 
adiabatically switched from the source to the first Q-switch cavity and then switched into the 
waveguide containing a Quarter Wave Plates (QWP) 
which rotates $\ket{\pm} \rightarrow \ket{\sigma_{\pm}}$.  The 
second Q-switch cavity is then used to adiabatically load the photon into the module cavity 
which contains the atomic system with a differential coupling between $\ket{\sigma_{\pm}}$ 
photons.  The final Q-switch cavity is then used to out-couple the photon back into the waveguide 
mode once the interaction is complete, where a second QWP rotates 
$\ket{\sigma_{\pm}} \rightarrow \ket{\pm}$.
In both schematics, the atom/cavity qubit has appropriate 
laser control such that it can be initialized and read out.  The measurement result of the atom qubit 
(denoted via the green and red readout channels of the external control lasers)
determines which eigenstate of $X^{\otimes N}$ the exiting photon train is 
projected to. A green measurement outcome corresponds to the photons projected into a +1 eigenstate 
while a red measurement outcome corresponds to projecting to the -1 eigenstate.}
\end{figure*}

Multi-partite entanglement is the most important resource needed when attempting 
to perform quantum processing.  Entanglement 
forms the basis of quantum algorithms~\cite{shor,grover}, secure cryptographic 
protocols and secret sharing~\cite{E91,SS}, Heisenberg limited optical lithography~\cite{lith}, 
and even a generic resource for a universal quantum 
computer~\cite{cluster1,cluster2}.  However, it has proven to be a difficult challenge to efficiently 
generate useful multi-qubit entangled states that can be used as a resource for 
all these disparate applications.  
Here we illustrate the construction of the photonic module [Fig.~\ref{fig:Fig1}].
A single atom/cavity system which leads to 
an extremely versatile device that can be used as a static resource for preparing entangled 
photonic states for computation and/or communication quickly, and with complete determinism.

Multi-partite entanglement can be prepared in systems such as trapped ions~\cite{CZ95} and 
solid state qubits~\cite{Kane,SC1,SC2,NV1,NV2,NV3}, 
but as resources for quantum cryptography and communications they 
are problematic due to a high sensitivity to environmental decoherence, qubit immobility 
and the inevitable incorporation of 
quantum bus protocols~\cite{qubus,MRAP,IF} and/or fusion 
methods~\cite{EO4,EO1,EO2,EO3} to solve problems related to information transport. 
In contrast, photonic qubits are extremely easy to move and 
are robust against decoherence.  However, performing appropriate gate operations to 
prepare photonic entanglement is extremely difficult, with experimental 
implementations of optical computing generally utilizing down-converted sources~\cite{op,op1}.  
Recently, more viable methods for the preparation of photonic entanglement have been 
developed~\cite{optics1,optics2,fusion,fusion1,fusion3, fusion2} and utilized~\cite{op2,op3}
based on the measurement induced non-linearities proposed by Knill, Laflamme and 
Milburn~\cite{KLM01}.  However, each of these measurement based methods result in 
probabilistic interactions and generally require number-discriminated photon detection.
The result of which is that large multi-photon entangled states need to be probabilistically 
grown.  Probabilistic state preparation and slow photon-detection results in long
preparation times for large entangled multi-photon states, requiring
significant resources for photon storage and limiting the applicability of
on-chip devices as entanglement resources.  

Although utilizing a single atomic qubit to mediate the preparation of photonic 
entanglement is not a new concept~\cite{schon} the nature of the interaction 
exhibited by this scheme leads to a simple and versatile plug and play device.  
A single module, or multiple connected modules, can be constructed and
with classical routing prepares entanglement without the downsides of
probabilistic interactions or single photon detection.  
Atom/cavity mediated entanglement is well understood and 
there exists several schemes which can be adapted for use in the modules.  These include the 
cavity-assisted interaction proposed by Duan, Wang and Kimble~\cite{duan,duan2} who utilized 
a similar network to achieve gate based photonic quantum computation
and recent experimental schemes from Schuster {\em et al.}~\cite{S07} which
demonstrated a photon non-demolition interaction using a
Cooper-pair box qubit and microwave cavity photons.  

We describe photonic modules for which the natural operation allows for the
preparation of any $N$ photon entangled state which can be described via the
stabilizer formalism of Gottesman~\cite{gott}.  
As such, the internal construction of each module is independent of the state being prepared, no single (or multi) 
photon detection is needed 
and the coherence time required for the atomic qubit is limited only by the time required 
to measure specific stabilizers describing the state  
(which can be small, even for large, highly entangled, multi-photon states).  
Therefore, the module is a completely generic resource, which can be applied to a 
vast variety of quantum applications.   

The possible uses for these modules are extensive.  The ability to prepare any 
stabilizer state allows for the deterministic preparation of {\em any} geometric graph state, 
including states appropriate for optical cluster state computation~\cite{cluster1,cluster2,Nielsen}.   
Bell state analyzers and factories are useful resources for quantum 
cryptographic protocols~\cite{E91}, quantum dense coding~\cite{DC1,DC2}, purification protocols and 
quantum repeaters~\cite{repeat,repeat2}.  Additionally, quick and deterministic preparation 
of $N$ photon GHZ states can be utilized in loss protection schemes for optical quantum 
computing~\cite{Ralph} and secret sharing protocols~\cite{SS}.  We begin the 
discussion with the the module shown in Fig.~\ref{fig:Fig1}a.  

The basic operation of the module is best understood if we choose to use it as a factory 
for two photon Bell states, defined through polarization as, 
\begin{eqnarray}
\ket{\Phi^+} = \frac{\ket{H}\ket{H} + \ket{V}\ket{V}}{\sqrt{2}}.
\end{eqnarray}
Given an appropriate single photon source, which 
can produce a train of single photon pulses of known polarization, 
separated by an interval $\Delta t$, a two photon train is prepared in the product state $\ket{H}_2^I\ket{H}_1^I$, 
and sequentially sent through the module.  The indices, $\{1,2\} = \{0, \Delta t\}$, 
represent the temporal mode of each single photon pulse, $I$ the spatial mode 
(in this case the optical input) and $\Delta t$
is predefined and must be greater than the total time a single photon is 
present within the network).

For a single photon passing through the module, the natural operation of the module, 
$M$, is given by,
\begin{equation}
\begin{aligned}
M\ket{+}^I\ket{\phi} \rightarrow \ket{+}^{O}\ket{\phi}, \\
M\ket{-}^I\ket{\phi} \rightarrow \ket{-}^{O}\ket{\phi'}.
\end{aligned}
\label{eq:trans1}
\end{equation}
Where $\ket{\pm} = (\ket{H}\pm \ket{V})/\sqrt{2}$, $\ket{\phi} = \alpha\ket{0}+\beta\ket{1}$ is the 
state of the atomic qubit, $\ket{\phi'} = X\ket{\phi} = \alpha\ket{1}+\beta\ket{0}$ and the 
indices $\{I,O\}$ represent the input and output optical modes. 

The atom/cavity system is positioned such that the cavity mode is coupled to the spatial mode
$\ket{\circ}^{B2}$, where $\circ$ denotes the photon polarization 
and cavity Q-switching (which allows for the adiabatic loading of a single photon into a cavity) 
is employed before and after the atom/photon interaction 
to ensure appropriate in- and out-coupling to and from the cavity. 
The mode $\ket{\circ}^{B1}$ contains an optical delay equal to the time required for the photon/atom 
interaction which is specific.  A single photon passing through the atom/cavity system 
must induce a photonic non-demolition bit-flip on the two-level 
atom, releasing the photon again into $\ket{\circ}^{B2}$ once the interaction is complete.  

If the photonic state is $\ket{+}$, the initial Half Wave Plate (HWP) will 
rotate the state to $\ket{H}$ after which it will continue into the mode $\ket{\circ}^{B1}$ and 
not interact with the atom.  The second Polarizing Beam Splitter (PBS) 
and HWP will then couple $\ket{\circ}^{B1}$ to the 
output mode and rotate $\ket{H}$ back to $\ket{+}$.  If the initial photonic state is $\ket{-}$, the 
HWP will rotate the state to $\ket{V}$ and the PBS will reflect the photon into the $\ket{\circ}^{B2}$ 
mode, where it flips the state of the atomic qubit.  The photon is 
then released back into $\ket{\circ}^{B2}$ where the second PBS and HWP will reflect the 
photon into the output mode and rotate it from $\ket{V}$ to $\ket{-}$. 
Therefore, the two basis states, $\ket{\pm}$, of a single photon passing through the 
module will enact the transformation $M$ shown in Eq.~\ref{eq:trans1}. 

For a two photon train, polarized in the state 
$\ket{H}_2^I\ket{H}_1^I = (\ket{+}_2^I+\ket{-}_2^I)(\ket{+}_1^I+\ket{-}_1^I)/2$, 
we are able to enact the same transformations on the photon/atom interaction, giving, 
\begin{equation}
\begin{aligned}
M_{2,1}\ket{H}_2^I\ket{H}_1^I \ket{\phi} =
&\frac{1}{\sqrt{2}}M_2\left(\ket{H}_2^I\ket{+}_1^O\ket{\phi} + \ket{H}_2^I\ket{-}_1^O\ket{\phi'}\right) \\
= &\frac{1}{2}\left(\ket{-}_2^O\ket{+}_1^O+\ket{+}_2^O\ket{-}_1^O\right)\ket{\phi'} \\
+ &\frac{1}{2}\ket{-}_2^O\ket{-}_1^O\ket{\phi''} + \frac{1}{2}\ket{+}_2^O\ket{+}_1^O
\ket{\phi},
\end{aligned}
\end{equation}
where $\ket{\phi''} = X\ket{\phi'} = X^2\ket{\phi}$.  Since $X^2 = I$, we can expand out 
the $\ket{\pm}$ states to give,
\begin{equation}
\begin{aligned}
M_{2,1}\ket{H}_2^I\ket{H}_1^I \ket{\phi} =
&\frac{1}{2}\left(\ket{H}_2^O\ket{H}_1^O+\ket{V}_2^O\ket{V}_1^O\right)\ket{\phi} \\
+&\frac{1}{2}\left(\ket{H}_2^O\ket{H}_1^O-\ket{V}_2^O\ket{V}_1^O\right)\ket{\phi'}. 
\end{aligned}
\end{equation}

After both photons have passed through the module the final step is to measure the state of 
the atom/cavity qubit.  If prior to the interactions, the atomic qubit is initialized in the 
state $\ket{\phi} = \ket{0}$ and the subsequent measurement is also $\ket{0}$, 
the photons are projected to the state,
\begin{equation}
\frac{1}{\sqrt{2}} (\ket{H}_2^{O}\ket{H}_1^{O} + \ket{V}_2^{O}\ket{V}_1^{O}),
\end{equation}  
which represents an even parity Bell state.  If the atom is 
measured in the state $\ket{1}$, the photons are projected to,
\begin{equation}
\frac{1}{\sqrt{2}}(\ket{H}_2^{O}\ket{H}_1^{O} - \ket{V}_2^{O}\ket{V}_1^{O}),
\end{equation}  
which is an odd parity Bell state.  The output pulse consists of the original two photon train which 
is now polarization entangled into a two photon Bell state.  Unlike other schemes, the measurement result of the 
atom/cavity system never collapses the photons to un-entangled states.  In fact, since the 
odd and even parity Bell states differ through local phase flips, either result is 
acceptable and a positive parity state can be prepared by applying a local, 
classically controlled phase flip on any photon once the atom/cavity qubit is measured.  

The preparation of the Bell state 
is therefore completely deterministic, with the classical result only giving parity information of 
the photon state.  Additionally, since positive and negative parity states are interchangeable 
through local Clifford gates, correction can be fed forward to the end of subsequent operations 
on the photonic state.
Although we explicitly considered the case when the induced 
operation was a bit-flip, any Hermitian unitary operation, $U^2 = I$, 
is acceptable provided it transforms the state 
of the atomic qubit between two orthogonal states. 

The transformation, $M$, shown in Eq.~\ref{eq:trans1} is also exhibited by the module 
illustrated in Fig.~\ref{fig:Fig1}b for a polarization dependent interaction.  
For appropriate atomic systems it is well known that there exists states with a differential 
dipole coupling between $\sigma_+$ and $\sigma_-$ polarized photons, e.g. NV$^-$ 
diamond~\cite{san} 
or Rubidium.  If the atomic 
coupling to the cavity mode is chosen such that only $\sigma_-$ polarized photons interacts 
with the atomic qubit, we are able to eliminate the interferometer shown in 
Fig.~\ref{fig:Fig1}a.  Instead, single 
photon wave plates are used to rotate $\ket{\pm} \leftrightarrow \ket{\sigma_\pm}$ before and 
after the atom/cavity interaction.  As the atom/photon interaction is polarization dependent, the 
transformation, $M$, for this modified version of the module still hold.  

The engineering of the 
module when a polarization dependent interaction is available is beneficial.  The 
lack of the interferometer implies that this structure can be directly fabricated in systems such 
as photonic bandgap crystals, with cavity Q-switching protocols fabricated on-chip to control the 
in- and out-coupling of the single photon pulses [Fig.~\ref{fig:Fig1}b].

\section{Atom/cavity interaction with photon pulse}
\label{sec:meth1}

The required atom/cavity interaction, crucial to the operation of the modules, has already been 
demonstrated at microwave frequencies.  
Schuster {\em et. al.}~\cite{S07} has demonstrated the 
non-destructive interaction we require where a single microwave 
cavity photon produces an effective stark shift on a classically driven transition on a cooper-pair box.  
These results were presented from the perspective of individual photon number detection, but the 
scheme can be inverted and used as the primary resource for a microwave version 
of the photonic module.  

In the optical regime, we can consider two separate schemes.  The proposal of Duan, Wang and 
Kimble~\cite{duan,duan2} considers a single atom/cavity system, such that a single photon 
reflecting from the cavity will produce a $\pi$ phase shift {\em only} if the atomic qubit is in the 
state $\ket{0}$.  Hence if the atomic qubit is 
initially placed in the state $\ket{+} = (\ket{0}+\ket{1})/\sqrt{2}$, subsequent photonic 
reflections will cause the atom to oscillate between the $\{\ket{+},\ket{-}\}$ states.  
Performing readout in the $\{\ket{0},\ket{1}\}$ basis corresponds to the 
(photonic) non-demolition, $X$ operation on the atom, as required.  

\begin{figure*}[tb!]
\includegraphics[width=1.6\columnwidth,clip]{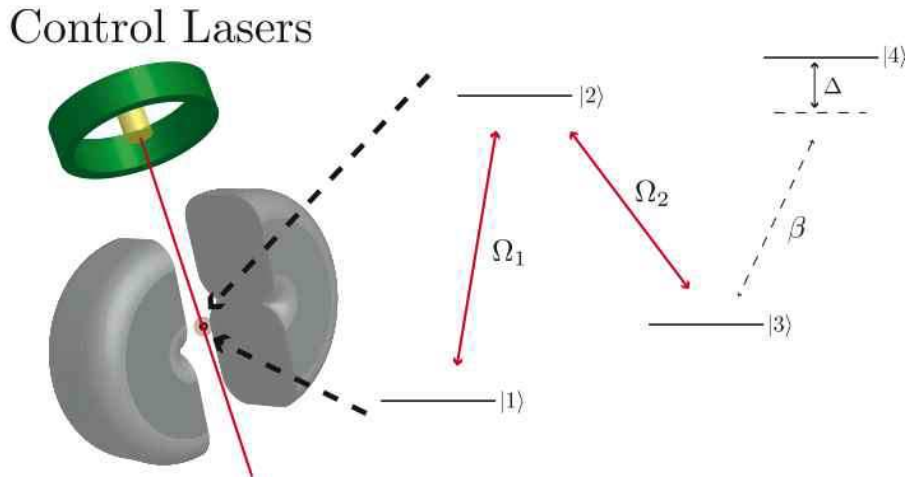}
\caption{\label{fig:Fig2} (Colour Online) 
\textbf{Four level atomic system required for the photonic module.}  
The atomic system 
is initialized in the $\ket{1}$ state and classical pumping fields are used to take $\ket{1}\rightarrow 
(\ket{1}-\ket{3})/\sqrt{2}$.  The single photon pulse is introduced to the cavity in a controlled manner 
using Q-switched cavities. The photon will induces a light shift on state 
$\ket{3}$ through the transition $\ket{3} \leftrightarrow \ket{4}$ (which 
may be polarization dependent), with strength $-\beta^2/\Delta$.  
Provided the photon 
remains in the cavity long enough, a $\pi$ phase shift can be induced on $\ket{3}$ without 
destroying the photon, taking $(\ket{1}-\ket{3})/\sqrt{2} \rightarrow 
(\ket{1}+\ket{3})/\sqrt{2}$.  After all atom/photon interactions have occurred, the classical fields are 
again applied and the atom read out in the $\{\ket{1},\ket{3}\}$ basis.}
\end{figure*}

A second method employs a four-level atom in the $N$ configuration, shown in Fig. \ref{fig:Fig2} which 
can be utilized for both the polarization dependent and independent modules 
illustrated in Fig.~\ref{fig:Fig1}.    
The general principal is to induce a phase shift $Z \equiv \sigma_z$ on the atom, 
conditional on the presence or absence of a photon in the cavity mode.  

{\em Initialization:}  The atomic system is initialized in the ground state, classical 
fields ($\Omega_1$, $\Omega_2$) are then used to take
$\ket{1}\rightarrow (\ket{1}-\ket{3})/\sqrt{2}$. 

{\em Interaction:}  We employ the idea of cavity Q-switching~\cite{Andy} in order to control the input/output 
pulse into the cavity system.  A single photon is adiabatically switched into the cavity, where it is 
off-resonant with the $\ket{3}\rightarrow \ket{4}$ transition (which may be polarization dependent),  
inducing a light-shift on the state $\ket{3}$.  The magnitude of the shift is well known and is given by,
$\delta = -\frac{\beta^2}{\Delta}$, for $\Delta \gg \beta$.  Therefore, to induce a phase shift of $\gamma$, the photon must be present in 
the cavity for a time given by, $t = \frac{\gamma\Delta}{\beta^2}$.
Consistent with the analysis in~\cite{Andy}, this implies that the photon storage time, $\kappa = 1/t$ must be 
$\kappa = \frac{\beta^2}{\gamma\Delta}$.  Taking $\gamma = \pi$ ensures that the the state   
$(\ket{1}-\ket{3})/\sqrt{2} \leftrightarrow (\ket{1}+\ket{3})\sqrt{2}$, performing the required $Z$ operation 
on the atomic system.  The photon is then Q-switched out of the cavity back into the optical mode
using appropriate shaping techniques~\cite{Andy,gerard}. 

{\em Readout:}  Readout is achieved by performing a second  transform on the atomic state 
using the classical fields, $\Omega_1$ and $\Omega_2$, in the same way as 
for initialization.  This takes the state 
$(\ket{1}-\ket{3})/\sqrt{2} \rightarrow \ket{1}$ and 
$(\ket{1}+\ket{3})/\sqrt{2} \rightarrow \ket{3}$.  Computational readout can be performed by classically 
pumping the transition $\ket{3} \leftrightarrow \ket{4}$ and observing photo-luminescence.  
Although the atom/photon interaction induces a $Z$ gate on the basis states 
$\{(\ket{1}\pm\ket{3})/\sqrt{2}\}$, the atomic rotations and readout in the 
$\{\ket{1},\ket{3}\}$ basis ensures that a bit-flip is performed on the atomic state.

{\em Operational Time:}  For this specific method, we can examine the transition 
time for a single photon in the module and compare our cavity requirements with systems currently 
in existence.  In general, we wish to maintain single-photon absorption probabilities on the 
$\ket{3} \leftrightarrow \ket{4}$ transition of less that $\zeta \ll 1$, hence 
$\Delta \geq \beta / \sqrt{\zeta}$.  For a $\pi$ phase shift, and choosing the equality, this 
corresponds to,
$t = (\kappa)^{-1} = \frac{\pi}{\beta \sqrt{\zeta}}$.
For a proof of concept device, we assume $\zeta = 0.1$, which corresponds to an average of one in ten photons 
being absorbed.  Consequently, we can examine $t$ as a function of some of the current experimental 
values for $\beta$ and $\kappa$ [Tab.~\ref{tab:cavity}],
\begin{table}[ht!]
\begin{center}
\vspace*{4pt}   
\begin{tabular}{c|c|c|c|c|c}
Cavity & $\beta$ (MHz) & $t$ at $\zeta = 10^{-1}$ & Exp. $t = \kappa^{-1}$ & $\Gamma$ (MHz) \\
\hline
Cs~\cite{Kimble} & $34$ & 0.29$\mu$s & 0.24$\mu$s & 2.6\\
Rb~\cite{Hinds} & $366$ & 27ns & 1.7ns & 6.3\\ 
NV$^-$~\cite{NORI} & $\approx 10^4$ & $\approx$ 1ns & $\approx$ 3.4ns & 83
\end{tabular}
\vspace*{4pt}
\caption{Estimates on Cavity requirements for various systems.  The Cavity from 
Boozer {\em et. al.}~\cite{Kimble} 
has been experimentally demonstrated, while the atom chip cavity of 
Trupke {\em et. al.}~\cite{Hinds} and 
the photonic bandgap cavity of Song {\em et. al.}~\cite{NORI} have yet to couple the atomic qubit.  
Hence we use their theoretical estimates for the coupling, $\beta$, and 
atomic decay rates, $\Gamma$, for Rubidium and NV$^-$~\cite{Andy,NV2} qubits.
The first column quotes the atom/cavity coupling while the second column estimates the 
required photon storage time in the cavity to invoke a $\pi$ phase shift in the atom with a single 
photon absorption probability of 10$\%$.  The final column quotes the current photon storage time 
which has been experimentally demonstrated (estimated) 
for each cavity system.  For both the 
Boozer and Trupke cavities, approximately an order of magnitude improvement in either the coupling 
constant or cavity lifetime is required.  Current estimates 
suggest that the photonic bandgap cavity will be able to exhibit the interaction with the 
fastest operational time of all the systems and is also more amenable to current cavity Q-switching 
protocols~\cite{Andy}.  The last column details estimates on atomic decay rates for the systems 
considered for each cavity, the ratio of the required photon storage time to the coherence 
time of the atomic system dictates the maximum Parity-weight a single module can measure 
in any one step.} 
\label{tab:cavity}
\end{center}
\end{table}  
The last column in Tab.~\ref{tab:cavity} specifies the atomic decay rates, $\Gamma$, 
for the atomic systems 
used in each cavity (Cesium, Rubidium and NV diamond).  The coherence time of the atomic 
system will dictate the maximum Parity-weight which can be measured in any one step using the 
module.  Hence taking the ratio of the required photon storage time to $\Gamma$, the Cs 
cavity of Boozer {\em et. al.}~\cite{Kimble} 
falls slightly short of being able to perform two-photon parity measurements (which 
are sufficient to prepare Bell states, GHZ states and linear cluster states~\cite{IF}).  The 
micro-cavity of Trupke {\em et. al.}~\cite{Hinds}, using Rb, theoretically has sufficient coherence to allow 
a parity check over six photons (sufficient for universal cluster states), while NV$^-$ in 
photonic bandgap cavities could allow for Parity-weights up to twelve~\cite{NORI, NV2}, allowing 
for a huge amount of flexibility in preparing highly entangled graph states very quickly.  

\section{Arbitrary entangled state preparation.}

The potential of these modules 
goes far beyond the preparation of Bell states.  
In fact, the unit can be augmented with appropriate single photon routing and local 
operations to prepare {\em any} entangled photon state that can be expressed in terms of 
stabilizers \cite{gott}.  These include,  codeword states for Quantum Error 
Correction, Bell states, GHZ states and arbitrary graph states
(of which cluster states are a specific topological subset).

A remarkable property of the module is that the number of 
entangled photons that are prepared depends only on the number sent through the 
module, no internal structure of the module needs to be altered to entangle more photons.  

To illustrate, consider an $N$ photon train, with each single 
photon pulse separated by $\Delta t$.  Each basis element, $|\psi\rangle$, of the 
state, $|\Psi\rangle$, can be written in the 
form, $\ket{\psi} = \bigotimes_{a=0}^{N-1} |c_a\rangle_{a} = 
\bigotimes_{a=0}^{N-1} [ |+\rangle_{a} + (-1)^{c_a} |-\rangle_{a} ]$, 
where $|c_a\rangle = \{|H\equiv 0\rangle, |V\equiv 1\rangle\}$ and 
$|\pm\rangle = (|H\rangle \pm |V\rangle)/\sqrt{2}$,
with each single photon pulse centered at time $t = a\Delta t$. 
As we have shown, the transformations performed by the module are given by, 
$M|+\rangle^{I}|\phi\rangle_{q} = |+\rangle^{O}|\phi\rangle_{q}$ and 
$M|-\rangle^{I}|\phi\rangle_{q} = |-\rangle^{O}|\phi'\rangle_{q}$, 
where, $\ket{\phi}_q$ is the state of the atomic qubit, $\ket{\phi'}_q = X\ket{\phi}_q$, 
$I$ and $O$ are the input/output modes of 
the module and, for clarity, we have omitted the time index for the pulse.
Assuming that the atomic system is initialized in the state $\ket{\phi}_q = \ket{0}$, 
an arbitrary basis state of $|\Psi\rangle_N$ transforms as,
\begin{widetext}
\begin{equation}
\begin{aligned}
M_{N,..,1}|\psi\rangle|0\rangle_{q} &= M_{N,..,1}\bigotimes_{a=0}^{N-1}\bigg{[} |+\rangle + (-1)^{c_a} |-\rangle \bigg{]}\ket{0}_q \\
&= \quad |0\rangle_{q} \bigotimes_{a=0}^{N-1} \bigg{[}|+\rangle + (-1)^{c_a} |-\rangle\bigg{]}_{Ev|-\rangle} 
+ \quad &|1\rangle_{q} \bigotimes_{a=0}^{N-1} \bigg{[}|+\rangle + (-1)^{c_a} |-\rangle\bigg{]}_{Od|-\rangle},
\end{aligned}
\end{equation}
\end{widetext}
where the first term represents all states of the tensor product formed with an even number of 
$|-\rangle$ states and the second term represents all tensor products formed with an odd number of 
$|-\rangle$ states. 
The even and odd components of the basis terms, $|\psi\rangle$, can be written in the 
following way,
\begin{widetext}
\begin{equation}
\begin{aligned}
 \bigotimes_{a=0}^{N-1}\bigg{[} |+\rangle + (-1)^{c_a} |-\rangle \bigg{]}_{Ev|-\rangle}
= \quad&\frac{1}{2}\bigotimes_{a=0}^{N-1}\bigg{[} |+\rangle + (-1)^{c_a} |-\rangle \bigg{]} + \quad
&\frac{1}{2}\bigotimes_{a=0}^{N-1}\bigg{[} |+\rangle + (-1)^{c_a+1} |-\rangle \bigg{]}, \\
 \bigotimes_{a=0}^{N-1}\bigg{[} |+\rangle + (-1)^{c_a} |-\rangle \bigg{]}_{Od|-\rangle}
= \quad &\frac{1}{2}\bigotimes_{a=0}^{N-1}\bigg{[} |+\rangle + (-1)^{c_a} |-\rangle \bigg{]} - \quad
&\frac{1}{2}\bigotimes_{a=0}^{N-1}\bigg{[} |+\rangle + (-1)^{c_a+1} |-\rangle \bigg{]}.
\end{aligned}
\end{equation}
\end{widetext}
Noting that the second term in each equation is simply the state 
$X^{\otimes N}|\psi\rangle$,
each basis term, $|\psi\rangle$, transforms as,
\begin{equation}
\begin{aligned}
M_{N,..,1}|\psi\rangle|0\rangle_{q} &= 
\frac{1}{2}(|\psi\rangle + X^{\otimes N}|\psi\rangle)|0\rangle_{q} \\ &+ \frac{1}{2}( 
|\psi\rangle -  X^{\otimes N}|\psi\rangle)|1\rangle_{q},
\end{aligned}
\end{equation}
and consequently, the total state, $\ket{\Psi} = \sum_j \beta_j \ket{\psi}_j $, transforms as,
\begin{equation}
\begin{aligned}
M_{N,..,1}|\Psi\rangle|0\rangle_{q} &=
\frac{1}{2}(|\Psi\rangle + X^{\otimes N}|\Psi\rangle)|0\rangle_{q} \\& + \frac{1}{2}( 
|\Psi\rangle -  X^{\otimes N}|\Psi\rangle)|1\rangle_{q}.
\end{aligned}
\end{equation}

Therefore, the natural operation of the module is to project the train of photons into a $\pm1$ eigenstate 
of the $X^{\otimes N}$ operator, i.e. any arbitrary $N$ photon state will be transformed 
to,
\begin{equation}
\begin{aligned}
M_{N,..,1}\ket{\Psi}_N\ket{0} = \quad &\frac{1}{2}\left(\ket{\Psi}_N+X^{\otimes N}\ket{\Psi}_N\right)
\ket{0} \\
+\quad &\frac{1}{2}\left(\ket{\Psi}_N-X^{\otimes N}\ket{\Psi}_N\right)\ket{1}.
\label{eq:op}
\end{aligned}
\end{equation} 
Where $\{\ket{0},\ket{1}\}$ are the states of the atom/cavity qubit and all $N$ photons have been 
passed through the module.  The measurement outcome of the 
atomic system will determine which eigenstate is projected, with local operations 
applied to switch between eigenstates.
The stabilizer formalism for describing large entangled states is extremely 
useful in this discussion as they are linked very closely to the concept of parity measurements. 

To prepare {\em any} $N$ photon stabilized state, a parity check is performed on the 
$N$ stabilizers which describe the state.  As each of the stabilizers for an arbitrary $N$ photon 
state are described via an $N$-fold tensor product of the operators $\{I,X,Y,Z\}$, the ability to 
perform a parity check of the operator $X^{\otimes N'}$ for $N' \leq N$ and apply local operations 
is sufficient to stabilize an arbitrary state with respect to any operator of this form.   
Therefore, if we assume that we can selectively route photons within the train (which is possible, as 
each pulse is temporally tagged) and 
apply local operations to any photon, the parity measurement performed by 
the module is sufficient to prepare any stabilizer state.

For a general $N$ photon state, $N$ parity checks are 
required.  This can either be done by constructing and utilizing $N$ separate modules, or 
it can be done by sequentially utilizing only one.  
If multiple modules are available, many parity checks can be done in parallel without 
waiting for atomic readout, potentially speeding up state preparation. 

As the stabiliser structure of the desired state dictates the number of photons passed through 
the module for each parity check, the coherence time of the atom/cavity system does not 
depend on the total number of photons in the entangled state.  Instead, the atomic system 
only has to maintain coherence until the parity of a specific stabiliser operator is measured,  
this is extremely beneficial.  
The number of non-Identity operators in any given stabilizer operator (which 
we denote the ``Parity-weight"), 
dictates the number of 
photons passed through the module in any one step and therefore the coherence time required 
for the atom/cavity system.  

For example, an $N$ photon cluster state appropriate 
for quantum computation, has a well known stabilizer structure~\cite{cluster1}, 
with a maximum Parity-weight of five.  Hence, regardless of the total size of the cluster, 
the atomic system only needs to maintain coherence long enough for five photons to pass through 
the module between initialization and measurement.  

Conversely, if the coherence time of the atomic system is 
short compared to $P_m\Delta t$, where $P_m$ is the maximum Parity-weight of the state and 
$\Delta t$ is the time required for a single photon to pass through the module, 
then fusion methods~\cite{Nielsen,EO1,EO3} can be employed to prepare 
states with large $P_m$.  For example, $N$ photon GHZ states have $P_m = N$,
corresponding to the stabilizer, $K = X^{\otimes N}$.  If the coherence time of the atomic system 
only allows for $N' < N$-dimensional parity checks to be performed at any one time, then multiple 
$N'$-GHZ states can be prepared and fused together via two-photon $ZZ$ parity measurements. 

\section{Conclusions}
We have detailed the construction of a photonic module which, given a steady source 
of single photons, can deterministically prepare a large class of useful photonic 
entangled states.  The construction of each module is generic and independent of 
which entangled state is being prepared, and the stabiliser nature of the entangled states 
implies that the coherence time of the atomic system only needs to be long compared with the 
maximum Parity-weight ($P_m$) $\times$ pulse separation ($\Delta t$) 
of the desired state
(which can be small, even for large multi-photon entangled states).  

The practical uses of these modules is quite extensive.  Multi-photon entangled states 
can be utilized for quantum computation, quantum cryptography, quantum dense coding, 
and quantum repeaters.

As the internal design of the photonic module is completely independent of the 
state being prepared, multiple modules, combined with an appropriate single 
photon source, optical wave plates and classical routing can be used to construct a static, on chip system, 
tailored for fast preparation of specific 
entangled states.  For example, pumping out large cluster states for computation, 
or multiple Bell pairs, in succession, for communications and cryptography.  

The engineering of an appropriate atom/photon interaction is still something that needs to be 
experimentally investigated.  
Cavity experiments reported in~\cite{Kimble,Hinds,NORI,S04,S07} 
show exceptional promise at both optical and microwave frequencies.
Ideally, once the required interaction has been experimentally demonstrated, additional 
engineering to realize the photonic module 
should be straight-forward. 

\section{Acknowledgments}
The authors thank 
J.H. Cole, Z. Evans, D. Jamieson, R. van Meter, G.J. Milburn, K. Nemoto, D.K.L. Oi, 
S. Prawer, R. Scholten, A.M. Stephens and C.-H. Su
for helpful discussions.  This work was supported, in part, by the Australian 
Research Council, the Australian Government, the 
US National Security Agency (NSA), Advanced Research 
and Development Activity (ARDA), Army Re- 
search Office (ARO) under Contracts Nos. W911NF-04- 
1-0290 and W911NF-05-1-0284, QAP and MEXT.


\end{document}